# Stellar Kinematics of the Milky Way Galaxy in the Post-Newton Gravity

Shubhen BISWAS[ł]

**Abstract:**

In this paper the recent interaction model over post-Newtonian theory of gravity has been used to study the stellar dynamics. It implies space time fabric that built over the massive object gets perturbed or literally contracted in presence of the test body having significant mass. In usual practice of general relativity the effect of interaction is missing or negligible but taking in consideration of this additional interaction the geodesic path followed by the test body must be deviated. From the celestial dynamics we know stellar rotation in the galaxy is not in accordance to Newton gravity and it prompts us to reconcile this interactive modified field for the galactic stars. Here we are to study the galactic stellar kinematics over introducing this additional perturbation. The galactic disk including the central super massive black hole and the star rotating around are being modelled theoretically as an effective two body system. In fact application of this model in the Milky Way shows the galactocentric solar rotational speed 220 km/s and is a good agreement with the observations.

ł Université de Tours, Parc de Grandmont, 37200 Tours, France.

Email: - shubhen3@gmail.com

## 1. Introduction:

The curiosity is a still overwhelming issue regarding the flat rotation curve for distant galactic stars after the discovery of Vera Rubin et al, **[1, 2].** In this occasion possible candidates are dark matter **[3]** and its immediate counter hypothesis in an empirical level the Modified Newtonian Dynamics (MOND) suggested by Milgrom **[4]** but it is not in accordance with the relativistic theory**[5]**. Following next the urge to replace the MOND several post Newtonian concept have been made to modify gravity **[6-8]**. The tensor-vector-scalar theory proposed by Bekenstein **[8]** even has some shortcomings to accounts the effect as explained by Sanders **[9]**. In getting dynamic solution for a system the best suited way is to construct a Lagrangian in post Newtonian (PN) level, to serve this purpose further work **[10-14]** have been done over Einstein Hilbert action principle. But all of them lack by the interaction effect among the mass sources.

In the following we would be happy to reconcile a two body system in a different manner over post Newton gravity introducing an additional correction term in spacetime perturbation regarding interaction **[15]** between them. The effect for two interacting bodies is described over the total perturbation of space time fabric. These will be as illustrated in accordance with the interactive model of post-Newton gravity by recent work **[15].** In post-Newton formalism every celestial body follows the geodesic path in the background of spacetime fabric manifested over the massive sources. The geodesic motion is connected through the perturbation of spacetime metric. To study the kinematics of a compact stellar body in the galactic spacetime fabric the total perturbation combined with interaction effect must be accounted for. Taken into considerations in case of the galactic spacetime it is merely sourced from the galactic disk containing billions of stars and the central super massive black hole (SMBH). It follows the stellar dynamics based on an effective two body system one is the galactic disk with the central SMBH and other is the rotating stellar body itself. The following sections are to evaluating the total perturbation and it's connected gravitational driving force field to formularize the celestial kinematics skipping the Lagrangian approach.

## 2. Geodesic motion in nearly flat spacetime:

The geodesic motion as like a free fall of a particle obeying least action principle **[16-18]** is

$$\frac{d^2x^\mu}{d\tau^2}+\Gamma^\mu_{\sigma\nu}\frac{dx^\sigma}{d\tau}\frac{dx^\nu}{d\tau}=0 \qquad (1)$$

Interpreting Christoffel connections that built over first order coordinate derivatives of the metric components

$$\Gamma^\mu_{\sigma\nu}=\frac{1}{2}g^{\mu\lambda}\left(\frac{\partial g_{\lambda\nu}}{\partial x^\sigma}+\frac{\partial g_{\lambda\sigma}}{\partial x^\nu}-\frac{\partial g_{\sigma\nu}}{\partial x^\lambda}\right) \qquad (2)$$

The geodesic path is parameterized over the proper time $\tau$, and here $\{x^\mu\}$ are stands for the coordinates $\left(x^0=ct,\{x^i\}=x^{1,2,3}\right)$ of the moving particle.

For a very slowly rotating with vast matter distribution the galactic spacetime fabric can be described over a static isotropic metric as given by Schwarzschild **[15-18]**.

$$ds^2{}_{Sch.}=-\left(1-\frac{2GM}{c^2r}\right)c^2dt^2+\left(1-\frac{2GM}{c^2r}\right)^{-1}dr^2+r^2d\theta^2+r^2sin^2\theta d\varphi^2 \qquad (3)$$

The PPN considerations over ignoring all the higher order terms of perturbations the Schwarzschild metric turns into quasi Minkowskian coordinates **[16,17],** here in Cartesian coordinates $(dx^0)^2 = c^2dt^2$ and $dX^2 = \sum_{i=1}^{3}\left(dx^i\right)^2$

$$ds^2{}_{Sch.} = -\left[1 - \frac{2GM}{c^2r}\right](dx^0)^2 + \left[1 + \frac{2GM}{c^2r}\right]\mathrm{d}X^2 \qquad (4)$$

The Minkowski flat spacetime metric,

$$\eta_{\mu\nu} = \begin{bmatrix} -1 & 0 & 0 & 0 \\ 0 & 1 & 0 & 0 \\ 0 & 0 & 1 & 0 \\ 0 & 0 & 0 & 1 \end{bmatrix} \qquad (5)$$

$$\left[\mathrm{g}_{\mu\nu}\right]_{Sch.} \approx \left[\eta_{\mu\nu}\right] + \left[\mathrm{h}_{\mu\nu}\right] = \begin{bmatrix} -1+\mathscr{b} & 0 & 0 & 0 \\ 0 & 1+\mathscr{b} & 0 & 0 \\ 0 & 0 & 1+\mathscr{b} & 0 \\ 0 & 0 & 0 & 1+\mathscr{b} \end{bmatrix} \qquad (7)$$

Comparing with the Minkowski flat spacetime metric, the Schwarzschild space-time metric in Eq.(7) has perturbations,

$$\mathscr{b} = \frac{2GM}{c^2r} = -\frac{2\Phi}{c^2} \ll 1 \qquad (6)$$

For nearly flat space when the gravity is very weak as likely at the outskirt galactic field the perturbations $\mathrm{h}_{\alpha\alpha} = \mathscr{b}$ and the inverse of the diagonal matrix in Eq.(7)

$$\left[\mathrm{g}^{\mu\nu}\right]_{Sch.} = \begin{bmatrix} -\left(\frac{1}{1-\mathscr{b}}\right) & 0 & 0 & 0 \\ 0 & \left(\frac{1}{1+\mathscr{b}}\right) & 0 & 0 \\ 0 & 0 & \left(\frac{1}{1+\mathscr{b}}\right) & 0 \\ 0 & 0 & 0 & \left(\frac{1}{1+\mathscr{b}}\right) \end{bmatrix} \approx \begin{bmatrix} -1-\mathscr{b} & 0 & 0 & 0 \\ 0 & 1-\mathscr{b} & 0 & 0 \\ 0 & 0 & 1-\mathscr{b} & 0 \\ 0 & 0 & 0 & 1-\mathscr{b} \end{bmatrix} \qquad (8)$$

$$\left[\mathrm{g}^{\mu\nu}\right]_{Sch.} \approx \left[\eta^{\mu\nu}\right] - \left[\mathrm{h}^{\mu\nu}\right] \qquad (9)$$

The suitable choice for getting the equations of motion following the geodesic is

$$\frac{d^2x^\mu}{d\tau^2} + \Gamma_{00}^{\mu}\left(\frac{dx^0}{d\tau}\right)^2 = 0 \qquad (10)$$

$$\Gamma_{00}^{\mu} = \frac{1}{2}\mathrm{g}^{\mu\lambda}\left(\frac{\partial \mathrm{g}_{\lambda\nu}}{\partial x^0} + \frac{\partial \mathrm{g}_{\lambda\sigma}}{\partial x^0} - \frac{\partial \mathrm{g}_{\sigma\nu}}{\partial x^\lambda}\right) \qquad (11)$$

In static gravitational field the time derivative of the metric is zero, i.e. $\frac{\partial \mathrm{g}_{\lambda\nu}}{\partial x^0} = \frac{\partial \mathrm{g}_{\lambda\sigma}}{\partial x^0} = 0$. But in nearly flat spacetime the perturbation is negligible $\left|h^{\mu\lambda}\right| \ll \left|\eta^{\mu\lambda}\right|$, now rewriting Eq.(11) in a reduced form

$$\Gamma_{00}^{\mu} = -\frac{1}{2}\eta^{\mu\lambda}\frac{\partial \mathrm{g}_{00}}{\partial x^\lambda} + \frac{1}{2}h^{\mu\lambda}\frac{\partial \mathrm{g}_{00}}{\partial x^\lambda} \approx -\frac{1}{2}\eta^{\mu\lambda}\frac{\partial \mathrm{g}_{00}}{\partial x^\lambda} \qquad (12)$$

The geodesic Eq.(10) now turns into simplest form

$$\frac{d^2x^\mu}{d\tau^2} \approx \frac{1}{2}\eta^{\mu\lambda}\frac{\partial \mathrm{g}_{00}}{\partial x^\lambda}\left(\frac{dx^0}{d\tau}\right)^2 \qquad (13)$$

To represent Eq.(13) in familiar coordinate time $'t'$ multiplying both sides by $\left(\frac{d\tau}{dt}\right)^2$ we have

$$\frac{d^2x^\mu}{dt^2} \approx \frac{1}{2}\eta^{\mu\lambda}\frac{\partial g_{00}}{\partial x^\lambda}\left(\frac{dx^0}{dt}\right)^2 \qquad (14)$$

For setting temporal coordinate $x^\mu = x^0$ in geodesic Eq.(14)

$$\frac{d^2x^0}{dt^2} \approx \frac{1}{2}\left(\eta^{00}\frac{\partial g_{00}}{\partial x^0} + \eta^{0i}\frac{\partial g_{00}}{\partial x^i}\right)\left(\frac{dx^0}{dt}\right)^2 \qquad (15)$$

The static condition implies $\frac{\partial g_{00}}{\partial x^0} = 0$ and also all the cross term $\eta^{0i} = 0$ leaves $\Gamma^0_{00} \approx \frac{1}{2}\left(\eta^{00}\frac{\partial g_{00}}{\partial x^0} + \eta^{0i}\frac{\partial g_{00}}{\partial x^i}\right) = 0$ and Eq.(15) becomes trivial for getting solution.

The non vanishing equations of motion is for spatial coordinates $x^\mu = \{x^i\}$

$$\frac{d^2x^i}{dt^2} \approx \frac{c^2}{2}\eta^{ii}\frac{\partial g_{00}}{\partial x^i} \qquad (16)$$

$$\frac{d^2\boldsymbol{r}}{dt^2} \approx \frac{c^2}{2}\boldsymbol{\nabla}(\mathscr{h}) \qquad (17)$$

Putting the perturbation in terms of Newton potential from Eq.(6) the corresponding force field giving back the Newton's gravity formula!

$$\frac{d^2\boldsymbol{r}}{dt^2} = \mathbf{f}(\boldsymbol{r}) \approx \frac{c^2}{2}\boldsymbol{\nabla}(\mathscr{h}) = -\boldsymbol{\nabla}(\Phi) \qquad (18)$$

Thus the weak field metric perturbation and Newton gravitational potential both equivalently inter related through the equation

$$h_{00} = \mathscr{h} = -\frac{2\Phi}{c^2} \qquad (19)$$

Our intention is to study of the stellar dynamics based on an effective two body system one is the galactic disk with the central SMBH and other is the stellar body rotating around. Now for a two body system in the weak field approximation the combined effect of perturbations that results following the post-Newton formalism **[15]** is

$$\mathscr{h}(\boldsymbol{r}) = [\,\mathscr{h}_1(\boldsymbol{r}) + \mathscr{h}_2(\boldsymbol{r}) - \mathscr{h}_1(\boldsymbol{r})\mathscr{h}_2(\boldsymbol{r})] \qquad (20)$$

The Eq. (20) is quite distinct from our normal point of view for a two body system; it not only contains the first and second terms for usual perturbation but considers an additional effect in perturbation due to the interaction between two sources.

## 3. Spacetime perturbation for extended objects:

In concern of the interaction effect the non-Newtonian gravity contributes to classical gravitational force field in a little bit different way! The average perturbation of spacetime for interaction due to the matter energy of the extended object can be obtained by taking the perturbation$(\mathscr{h})$experienced at the two opposite surfaces. Let the two opposite surfaces of the compact body with a close distant apart from points $(\boldsymbol{r} - \boldsymbol{\delta r})$ and $(\boldsymbol{r} + \boldsymbol{\delta r})$as in **Fig.1** (a) have perturbations taking only first order derivative of Taylor series

$$\mathscr{b}(\boldsymbol{r} \pm \boldsymbol{\delta r}) \approx \mathscr{b}(\boldsymbol{r}) \pm \frac{\partial \mathscr{b}(\boldsymbol{r})}{\partial r} \delta r \qquad (21)$$

$$\mathscr{b}(\boldsymbol{r}) \approx \frac{1}{2}[\mathscr{b}(\boldsymbol{r} - \boldsymbol{\delta r}) + \mathscr{b}(\boldsymbol{r} + \boldsymbol{\delta r})] \qquad (22)$$

In usual way the celestial motion is followed by the geodesic path in a space time that already manifested over the massive object $M$ . But this spacetime fabric gets further perturbation in putting a test body for its mass '$m$' at the point '$\boldsymbol{r}$' as **Fig.1** (b) due to interaction. Interaction between two sources of masses '$m$' and '$M$' literally shrinks the spacetime fabric around the mass $m$, actually increases the spacetime curvature near the surroundings of the test body and consequently we would have different geodesic path, deviated from the former.

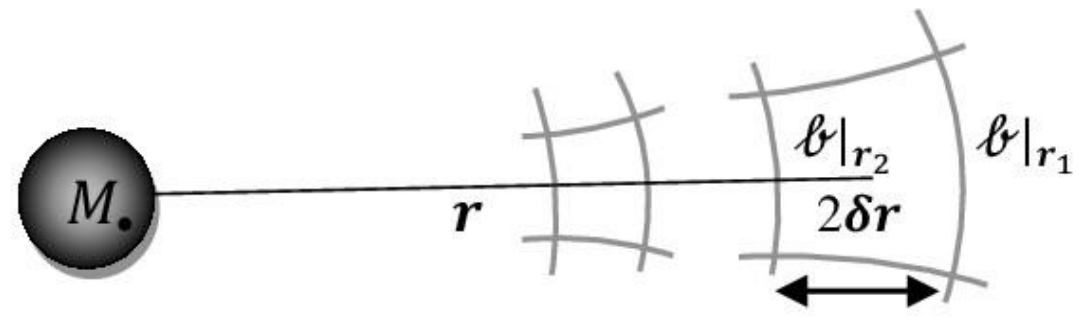

**Fig.1** (a) The Space-time fabric created only by the single source '$M$'

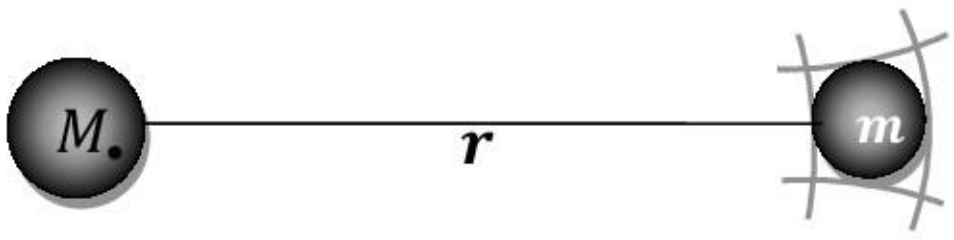

**Fig.1** (b) Perturbed Space-time fabric around the mass '$m$' over interaction

The interaction purturbation just discussed comes to play to induce a distinct force field which is not considered in pure Newtonian gravity! And in an extended nature this idea is derived from standard general relativistic approach of spacetime geometry **[15]**. Let us consider in a practical situation when a so called spherical massive body or a celestial object having significant size is embedded in the galactic disk. The space time fabric that builds over by the matter energy of the galactic disk comes to interact with the stellar spherical body and manifests an imbalance in perturbations between two opposite surfaces.

Generally in the case of geodesic motion we used to consider the path traced by the point mass '$m$' is the world line in the background of space time manifested by the mass $M \gg m$ . But in real situation the mass has significant physical size and shape hence the geodesic path of the physical body need to be described over the world tube. Here our choice of '$\delta r$' is the radius of the spherical body of mass '$m$' which is placed on the back ground of space-time fabric of the mass '$M$'. The geodesic path corresponds to the effective perturbation in Eq. (20) will now describe the stellar trajectory.

## 4. The kinematics of extended stellar body:

In the Eq. (20) the first and second terms signifies usual perturbations and last is the interaction effect. Thus in computing the rotation of the stellar body let us think galactic disk with SMBH as the first object and the second one is the celestial body itself separated by a distance '$\boldsymbol{r}$'.

For the galaxy SMBH plus disk mass **Fig.2**, taking galactic potential,

$$\varphi_{\mathcal{G}} = [\varphi_{\bullet} + \varphi_d] \qquad (23)$$

$$\mathscr{h}_1(\boldsymbol{r}) = -\frac{2\varphi_{\mathcal{G}}}{c^2} = -\frac{2(\varphi_{\bullet}+\varphi_d)}{c^2} = \frac{2G}{c^2}\left[\frac{M_{\bullet}+M_d(r)}{r}\right] \qquad (24)$$

The second one an extended object is a stellar body and self potential throughout the body is not same and we need to calculate the average. This can be done just by dividing the extended object in numbers of concentric spherical shells. At $r_0$ shell element faces potential due to outside and inside mass of the shell **[19]**,

$$\mathscr{h}_2(\boldsymbol{r}_0) = -\frac{2\varphi_0}{\mathrm{c}^2} = \frac{Gm}{\mathrm{c}^2 R^3}[3R^2 - {r_0}^2] \qquad (25)$$

Interaction potential at the shell,

$$\mathscr{h}_1(\boldsymbol{r})\mathscr{h}_2(\boldsymbol{r}_0) = \frac{4(\varphi_{\bullet}+\varphi_d)\varphi_0}{c^4} = \frac{2G^2 m}{c^4 R^3}\left[\frac{M_{\bullet}+M_d(r)}{r}\right][3R^2 - {r_0}^2] \qquad (26)$$

Using back Eq. (20) and implying Eqns. (24), (25) and (26) for each and every shell elements of the extended celestial body, the post-Newton spacetime perturbation at the rear side given as a function of $\boldsymbol{r}_1 = (\boldsymbol{r} + \boldsymbol{r}_0)$ outward to galactic centre is

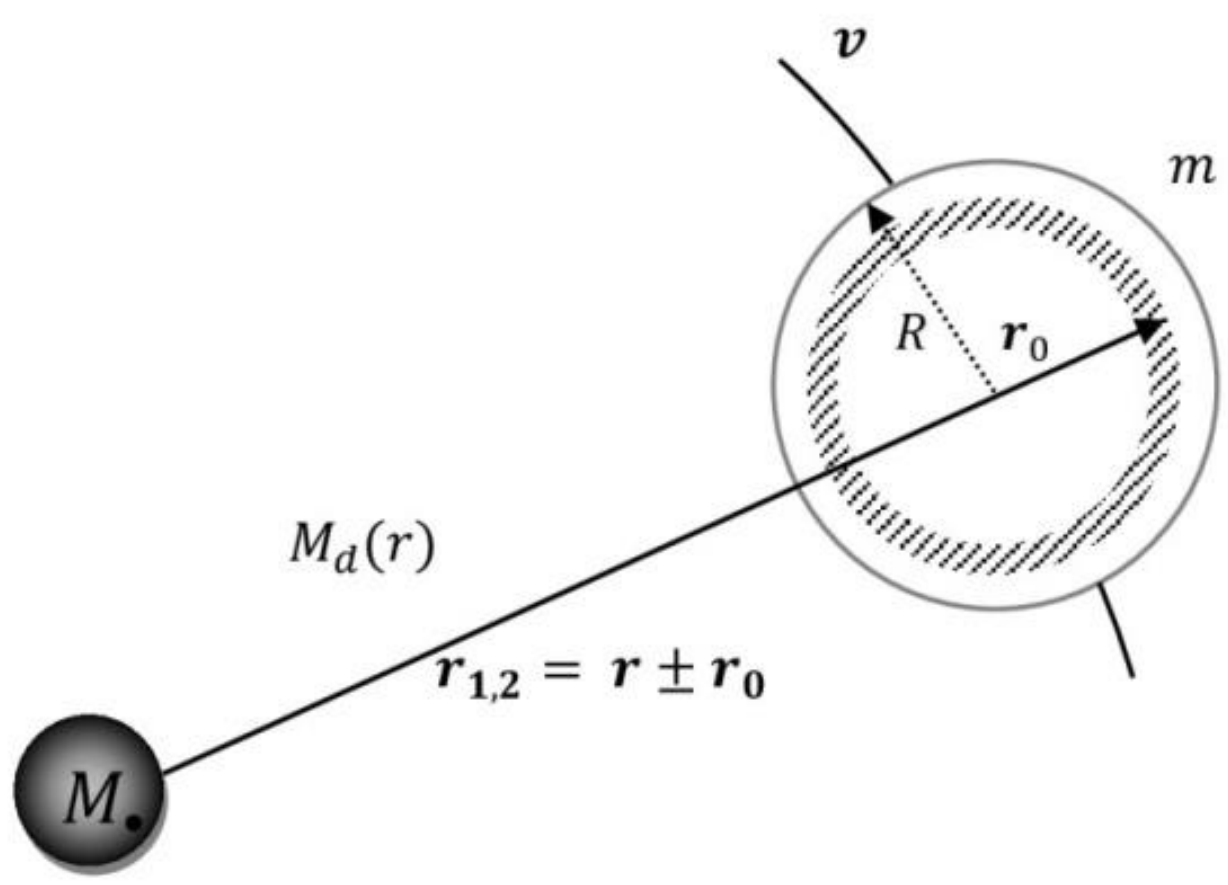

**Fig.2** Rotating star of mass $'m'$ around the galactic nuclei

The corresponding two body perturbation

$$\mathscr{h}(\boldsymbol{r}+\boldsymbol{r}_0) = \mathscr{h}_1(\boldsymbol{r}+\boldsymbol{r}_0) + \mathscr{h}_2(\boldsymbol{r}+\boldsymbol{r}_0) - \mathscr{h}_1(\boldsymbol{r}+\boldsymbol{r}_0)\mathscr{h}_2(\boldsymbol{r}+\boldsymbol{r}_0) \qquad (27)$$

The perturbation inside a spherical stellar body for its own matter energy is independent on $'\boldsymbol{r}'$

$$\mathscr{h}_2(\boldsymbol{r}+\boldsymbol{r}_0) = \frac{Gm}{\mathrm{c}^2 R^3}[3R^2 - {r_0}^2] = \mathscr{h}_2(\boldsymbol{r}_0) \qquad (28)$$

$$\mathscr{b}(\boldsymbol{r}+\boldsymbol{r}_0)=\mathscr{b}_1(\boldsymbol{r}+\boldsymbol{r}_0)+\mathscr{b}_2(\boldsymbol{r}_0)-\mathscr{b}_1(\boldsymbol{r}+\boldsymbol{r}_0)\mathscr{b}_2(\boldsymbol{r}_0) \qquad (29)$$

Just the opposite side close to the galactic centre, using Eq.(29) in similar way for this side $\boldsymbol{r}_2=\boldsymbol{r}-\boldsymbol{r}_0$ that facing the galactic centre, we have

$$\mathscr{b}(\boldsymbol{r}-\boldsymbol{r}_0)=\mathscr{b}_1(\boldsymbol{r}-\boldsymbol{r}_0)+\mathscr{b}_2(\boldsymbol{r}-\boldsymbol{r}_0)-\mathscr{b}_1(\boldsymbol{r}-\boldsymbol{r}_0)\mathscr{b}_2(\boldsymbol{r}-\boldsymbol{r}_0) \qquad (30)$$

Also likely Eq.(28)

$$\mathscr{b}_2(\boldsymbol{r}-\boldsymbol{r}_0)=\frac{Gm}{\mathrm{c}^2R^3}[3R^2-{r_0}^2]=\mathscr{b}_2(-\boldsymbol{r}_0) \qquad (31)$$

$$\mathscr{b}(\boldsymbol{r}-\boldsymbol{r}_0)=\mathscr{b}_1(\boldsymbol{r}-\boldsymbol{r}_0)+\mathscr{b}_2(-\boldsymbol{r}_0)-\mathscr{b}_1(\boldsymbol{r}-\boldsymbol{r}_0)\mathscr{b}_2(-\boldsymbol{r}_0) \qquad (32)$$

Taking the contributions from both sides of the celestial body by adding Eqns. (29) and (32)

$$[\mathscr{b}(\boldsymbol{r}+\boldsymbol{r}_0)+\mathscr{b}(\boldsymbol{r}-\boldsymbol{r}_0)]=[\mathscr{b}_1(\boldsymbol{r}+\boldsymbol{r}_0)+\mathscr{b}_1(\boldsymbol{r}-\boldsymbol{r}_0)]+[\mathscr{b}_2(\boldsymbol{r}_0)+\mathscr{b}_2(-\boldsymbol{r}_0)]-$$
$$[\mathscr{b}_1(\boldsymbol{r}+\boldsymbol{r}_0)\mathscr{b}_2(\boldsymbol{r}_0)+\mathscr{b}_1(\boldsymbol{r}-\boldsymbol{r}_0)\mathscr{b}_2(-\boldsymbol{r}_0)] \qquad (33)$$

But $\mathscr{b}_2(\boldsymbol{r}_0)=\mathscr{b}_2(-\boldsymbol{r}_0)$ and using condition of Eq.(22) the Eq.(33) gives the average perturbation

$$\mathscr{b}(\boldsymbol{r})=\mathscr{b}_1(\boldsymbol{r})+\mathscr{b}_2(\boldsymbol{r}_0)-\mathscr{b}_1(\boldsymbol{r})\mathscr{b}_2(\boldsymbol{r}_0) \qquad (34)$$

Putting in terms of potentials of Eqns. (24) to (26)

$$\mathscr{b}(\boldsymbol{r})=-\left[\frac{2\varphi_{\mathcal{G}}(\boldsymbol{r})}{c^2}+\frac{2\varphi_0(\boldsymbol{r}_0)}{c^2}+\frac{4\varphi_{\mathcal{G}}(\boldsymbol{r})\varphi_0(\boldsymbol{r}_0)}{c^4}\right] \qquad (35)$$

Using Eq.(35) the equation of motion for spatial coordinates in Eq.(17) turns into

$$\frac{d^2\boldsymbol{r}}{dt^2}\approx-\boldsymbol{\nabla}\left[\varphi_{\mathcal{G}}(\boldsymbol{r})+\varphi_0(\boldsymbol{r}_0)+\frac{2\varphi_{\mathcal{G}}(\boldsymbol{r})\varphi_0(\boldsymbol{r}_0)}{c^2}\right] \qquad (36)$$

$$\frac{d^2\boldsymbol{r}}{dt^2}\approx-\boldsymbol{\nabla}[\varphi_{\mathcal{G}}(\boldsymbol{r})]-\boldsymbol{\nabla}\varphi_0(\boldsymbol{r}_0)-\frac{2\boldsymbol{\nabla}[\varphi_{\mathcal{G}}(\boldsymbol{r})]\varphi_0(\boldsymbol{r}_0)}{\mathrm{c}^2}-\frac{2[\varphi_{\mathcal{G}}(\boldsymbol{r})]\boldsymbol{\nabla}\varphi_0(\boldsymbol{r}_0)}{\mathrm{c}^2} \qquad (37)$$

$$\frac{d^2\boldsymbol{r}}{dt^2}\approx-\boldsymbol{\nabla}[\varphi_{\mathcal{G}}(\boldsymbol{r})]\left(1+\frac{2\varphi_0}{\mathrm{c}^2}\right)-\boldsymbol{\nabla}\varphi_0(\boldsymbol{r}_0)-\frac{2[\varphi_{\mathcal{G}}(\boldsymbol{r})]\boldsymbol{\nabla}\varphi_0(\boldsymbol{r}_0)}{\mathrm{c}^2} \qquad (38)$$

Considering main sequence stars with very weak surface potentials, $\varphi_0 \ll \mathrm{c}^2$

$$\frac{d^2\boldsymbol{r}}{dt^2}\approx-\boldsymbol{\nabla}[\varphi_{\mathcal{G}}(\boldsymbol{r})]-\boldsymbol{\nabla}\varphi_0(\boldsymbol{r}_0)-\frac{2[\varphi_{\bullet}(\boldsymbol{r})+\varphi_d(\boldsymbol{r})]\boldsymbol{\nabla}\varphi_0(\boldsymbol{r}_0)}{\mathrm{c}^2} \qquad (39)$$

For outskirt the galactic disk mass profile is linear **[20]** that claims a constant disk potential $\varphi_d(\boldsymbol{r_1})\approx\varphi_d(\boldsymbol{r_2})\approx\varphi_d(\boldsymbol{r})\approx\alpha_d$, at least in a scale of stellar diameter. But the potential due to galactic nuclei is no more linear $\varphi_{\bullet}(\boldsymbol{r_1})\neq\varphi_{\bullet}(\boldsymbol{r_2})$, in case of our Milky Way galaxy centre there is a SMBH with minimum centralized mass $M_{\bullet}=4\times10^5M_{\odot}$ [**21, 22**], must be accompanied with inverse potential!

$$\frac{d^2\boldsymbol{r}}{dt^2}\approx-\boldsymbol{\nabla}[\varphi_{\mathcal{G}}(\boldsymbol{r})]-\frac{2[\varphi_{\bullet}(\boldsymbol{r})]\boldsymbol{\nabla}\varphi_0(\boldsymbol{r}_0)}{\mathrm{c}^2}-\boldsymbol{\nabla}\varphi_0(\boldsymbol{r}_0)\left[1+\frac{2\alpha_d}{\mathrm{c}^2}\right] \qquad (40)$$

Using **Fig:2** we have the condition

$${r_0}^2=|\boldsymbol{r_1}-\boldsymbol{r}|^2 \qquad (41)$$

$$\boldsymbol{\nabla}r_0=-\boldsymbol{\nabla}r=-\hat{\boldsymbol{r}} \qquad (42)$$

Taking compact object having constant density and using relation (42) in Eq.(25)

$$\boldsymbol{\nabla}\varphi_0(\boldsymbol{r}_0) = \frac{Gmr_0}{R^3}[\boldsymbol{\nabla}r_0] = \frac{Gmr_0}{R^3}[-\hat{\boldsymbol{r}}] = -\left(\frac{Gm_0}{{r_0}^2}\right)\hat{\boldsymbol{r}} \qquad (43)$$

For extended celestial object as a whole we need to compute the average force field to the total size and mass using Eq.(40) with considering Eq.(43)

$$\langle\frac{d^2r}{dt^2}\rangle = -\boldsymbol{\nabla}\left[\varphi_{\mathcal{G}}(\boldsymbol{r})\right] + \frac{2[\varphi_\bullet(r)]}{c^2}\langle\frac{Gm_0}{{r_0}^2}\rangle\hat{\boldsymbol{r}} + \langle\frac{Gm_0}{{r_0}^2}\rangle\left[1+\frac{2\alpha_d}{c^2}\right]\hat{\boldsymbol{r}} \qquad (44)$$

$$\langle\frac{Gm_0}{{r_0}^2}\rangle = \frac{\int_0^m\int_0^R \frac{Gm_0}{{r_0}^2}dm_0 dr_0}{\int_0^m dm_0\int_0^R dr_0} = \frac{3}{20}\frac{Gm}{R^2} \qquad (45)$$

In Eq.(45) $\frac{Gm}{R^2} = \mathrm{g}$ is star's surface gravity, now in post-Newton formalism the average force field to a celestial body in the galaxy using Eqn.(45) to the Eq.(44) gives

$$\langle\frac{d^2r}{dt^2}\rangle = -\boldsymbol{\nabla}\left[\varphi_{\mathcal{G}}(\boldsymbol{r})\right] + \frac{3[\varphi_\bullet(r)]\mathrm{g}}{10c^2}\hat{\boldsymbol{r}} + \frac{3\mathrm{g}}{20}\left[1+\frac{2\alpha_d}{c^2}\right]\hat{\boldsymbol{r}} \qquad (46)$$

The last square bracketed term in the right of Eq.(46) is a constant independent of radial distance$(\boldsymbol{r})$; we can discard it from the governing central force field.

$$\mathfrak{H}(\boldsymbol{r}) = -\boldsymbol{\nabla}\left[\varphi_{\mathcal{G}}(\boldsymbol{r})\right] + \frac{3[\varphi_\bullet(r)]\mathrm{g}}{10c^2}\hat{\boldsymbol{r}} \qquad (47)$$

The effective galactic field in the Eq.(47) excels the Newtonian model of gravity! For a celestial object it introduces an additive interaction term due to the galactic nuclei.

## 5. Solar rotational motion:

Milky Way home galaxy has estimated baryonic mass, $M_{\mathcal{G}} = 5.04\times10^{10}M_{\odot}$**[23]**, $M_{\odot} = 1.988\times10^{30}kg$[**24, 25**], Solar distance from galactic nuclei $r = 8.15\,kpc \approx 2.5\times10^{20}\ meter$ **[26]** and constants $G \approx 6.67\times10^{-11}SI$, $c = 3\times10^{8}\ meter/sec$ implies,

$$\left|\boldsymbol{\nabla}\left[\varphi_{\mathcal{G}}(\boldsymbol{r})\right]\right| = \frac{GM_{\mathcal{G}}}{r^2} \approx 0.107\times10^{-10}\ meter.sec^{-2} \qquad (48)$$

$$\mathrm{g}_{\odot} = \frac{GM_{\odot}}{{R_{\odot}}^2} = 2.74\times10^{2}\ meter.sec^{-2}[\mathbf{25}] \qquad (49)$$

$$|\varphi_\bullet(\boldsymbol{r})| = \frac{GM_\bullet}{r} \approx 2.12\times10^{5}\ meter^2.sec^{-2} \qquad (50)$$

And the interaction field

$$\frac{3|\varphi_\bullet(r)|\mathrm{g}}{10c^2} \approx 1.93\times10^{-10}\ meter.sec^{-2} \qquad (51)$$

The interaction field in Eq.(51)is an inverse field$\sim\frac{1}{r}$ and contributing 94.7% of the total field in Eq.(47) unlike for matter Newtonian galactic field just giving 5.3% only! Thus interaction effect can be thought of as the possible successor candidate of the dark matter. For governing the kinematics of the outskirt celestial body like our Sun only the one to one interaction of the stars to the SMBH binary system does matter.

Using the Newtonian mechanics the central force of Eq.(47) that bounds the outskirt celestial body to rotate in a orbit at the radial distance $'r'$ with speed $'v',$ we have the equation of motion,

$$\mathfrak{H}(\boldsymbol{r}) = \frac{v^2}{r} \approx \frac{3|\varphi_\bullet(\boldsymbol{r})|\mathrm{g}}{10\mathrm{c}^2} \qquad (52)$$

$$v = \left(\frac{3GM_\bullet \mathrm{g}}{10c^2}\right)^{\frac{1}{2}} \qquad (53)$$

Following Eq.(53) the orbital speed of a outskirt stellar body in the galaxy is independent of the distances$(\boldsymbol{r})$ as expected from the rotation curve analysis. But the speed varies as with the stellar surface gravitational field $'\mathrm{g}'$! The Eq.(53) results the galactocentric solar orbital speed $v_\odot \approx 220km/sec$!

## 6. Conclusions and remarks:

Though the galaxy itself is an N-body system but collective influence to spacetime at the outskirt star can be constructed taking the central SMBH together with the disk. This led us treating stellar dynamics as an effective two body formalism owing to Eq. (20). Now in the interaction model **[15]** the synergy between two gravitating bodies over the perturbed metric is quite reasonable describing the two body dynamics. Constructing a suitable Lagrangian out of the post Newtonian extended interaction potential model can be a good choice. But the alternative straightforward way to study the two body kinematics in case galactic star of the above preceding sections give quite satisfactory results! In getting directly the orbital speed of a outskirt celestial body following Eq.(47) the formula Eq.(53) gives the expected result as IAU recommended solar rotational speed $220\ km.sec^{-1}$**[26, 27]** and is quite relevant as far as recent observations **[28-32]**. Also the Eq.(47) shows that effective galactic field not only contains the Newtonian inverse square field$\sim \frac{1}{r^2}$ but additionally an inverse interaction central field$\sim\frac{1}{r}$. The existence of inverse field demands Bertrand's unbound orbits **[33]** and is responsible for the galactic whirling of the outskirt celestial objects. Now do we have any data that could justify any connection between stellar surface gravity to the radial velocity? Moschella et al **[34]** found positive correlation of *Gaia* radial velocities and the differences in surface gravity!

**Acknowledgements:** I am undoubtedly grateful to Prof. Stam NICOLIS for having direct suggestions to present this manuscript in its final shape.

**Data Availability:**
The data used to compute and study the rotational speed for the Sun.

**SBMBH mass,**
Milky Way galactic centre SMBH, a lower limit to the mass of Sgr A* of mass is $4 \times 10^5 M_\odot$
(Doeleman, S., Weintroub, J., Rogers, A.et al, 2008, **455**, 78-80), DOI:10.1038/nature07245.
(Roelofs, F. et al. 2019, A&A 625, A124), https://doi.org/10.1051/0004-6361/201732423

**The Solar distance from the galactic centre**, $\sim 10^{20}$ *meter,*
(Sch¨onrich, R., 2012, MNRAS, 14), https://arxiv.org/pdf/0902.3913.pdf

**Solar parameters**
https://nssdc.gsfc.nasa.gov/planetary/factsheet/sunfact.html

http://hyperphysics.phy-astr.gsu.edu/hbase/Solar/sun.html

**The rotational velocity of Sun around the galactic centre,**
https://www.nsf.gov/discoveries/disc_summ.jsp?cntn_id=114090&org=NSF,

(Schönrich, R., 2012, MNRAS, 14), https://arxiv.org/abs/1207.3079

(Reid, M. J. et al. 2009, Astrophys.J.700:137-148), https://arxiv.org/pdf/0902.3913.pdf

(Ghez, A. M., Salim, S., Weinberg, N. N., et al. 2008, ApJ, 689:1044)
https://arxiv.org/pdf/0808.2870v1.pdf
(Bovy. J., et al, 2012, ApJ 759,131) https://doi.org/10.1088/0004-637X/759/2/131

**Declarations:** For this work there is no Funding and/or Conflicts of interests/Competing interests.